\documentclass[reqno,11pt]{article}

\usepackage{amsmath,amsfonts,amsgen,amstext,amsbsy,
amsopn,amsthm, amssymb,latexsym,amssymb,amsthm}

\newtheorem{theorem}{Theorem}[section]
\newtheorem{lemma}[theorem]{Lemma}
\newtheorem{proposition}[theorem]{Proposition}
\newtheorem{corollary}[theorem]{Corollary}

\newtheorem{assumption}[theorem]{Assumption}

\newtheorem{claim}[theorem]{Claim}
\numberwithin{equation}{section}
\numberwithin{theorem}{section}

\newcommand{\mA}{{\bf A}}
\newcommand{\mAM}{{\bf A}_M}
\newcommand{\mB}{{\bf B}}
\newcommand{\mC}{{\bf C}}

\newcommand{\mG}{{\bf G}}

\newcommand{\mK}{{\bf K}}
\newcommand{\mT}{{\bf T}}
\newcommand{\smT}{\sqrt{\bf T}}

\newcommand{\mD}{{\bf D}}

\newcommand{\tmA}{\tilde{\bf A}}
\newcommand{\tmAM}{\tilde{\bf A}_M}

\def\bd{\begin{displaymath}}
\def\ed{\end{displaymath}}

\def\qed{\hbox{\hskip 6pt\vrule width6pt height7pt depth1pt
    \hskip1pt}\bigskip}

\def\runinend{\enspace}

\def\ackname{Acknowledgement\runinend}
\def\acknowledgements{\par\addvspace{16 pt}\rmfamily
\def\ackname{{\it Acknowledgements}\runinend}
\trivlist\if!\ackname!\item[]\else
\item[\hskip\labelsep
{\bf\ackname}]\fi}%
\def\endacknowledgements{\endtrivlist\addvspace{6pt}}

\begin{document}
\begin{center}{\bf \Large Stationary state solutions for a gently stochastic 
nonlinear wave equation with ultraviolet cutoffs}
\end{center}

    {\large Yao Wang}\footnotemark\footnotetext[1]{Electronic mail: yao@math.umass.edu}\\
\indent {\it Department of Mathematics, Department of Mathematics and Statistics,\\ 
\indent University of Massachusetts,Amherst, MA 01003}\\

  {\large Lawrence E. Thomas}\footnotemark\footnotetext[2]{Electronic mail: let@virginia.edu}\\
\indent    {\it Department of Mathematics, University 
of Virginia, Charlottesville, VA 22904}

 \begin{abstract} We consider a non-linear, one-dimensional wave
equation system with finite-dimensional stochastic driving terms  and with
weak dissipation. A stationary process that solves the system is used to model steady-state non-equilibrium heat flow through a non-linear medium.
We show existence and uniqueness of invariant measures for the system modified
with ultraviolet cutoffs, and we obtain estimates for the field
covariances with respect to these measures, estimates that are uniform in the cutoffs.
Finally, we discuss the limit of these measures as the ultraviolet cutoffs are removed.\\



\end{abstract}

\section{Introduction}
In this article, we consider a dynamical system of equations for a
scalar field in one space dimension with ultraviolet cutoff, where 
the equations have a non-linear
term, dissipative terms, and stochastic driving terms. 
We will be concerned with the
 problems of constructing measures on the space of
field configurations invariant under the time evolution of
the system, and of estimating the covariance of the field with respect to
these measures. 

  This system of equations  provides a model for heat flow; the
  system is a variant of models introduced and examined in detail by
  Eckmann, Pillet, and Rey-Bellet (\cite{EPR1,EPR2}, see also
  \cite{RT1,RT2,RT3, EH1,RB}).  These authors considered a finite
  system of  non-linear oscillators coupled to two (or more)
  free fields that are governed by linear wave equations.  The free fields,
  which model heat reservoirs, are given Gaussian-distributed random
  initial conditions (Gibbs states), but in general at different 
  temperatures $T_i$ for each of the free fields.  The equations of
  motion for the free fields are readily integrated, leaving a system
  of stochastic equations for the oscillators.  For suitable
  couplings, these authors found that the equations were Markovian
  stochastic differential equations.  The equations given below, then,
  are simply the analogue of these stochastic differential
  equations, but with the oscillators replaced by a non-linear scalar
  field in one space dimension with bounded support.  We examine these
  equations with an ultraviolet cutoff, i.e. where Fourier modes of
  the field are set to zero for mode numbers $\{n\}$, $n >M$, and we
  consider the limiting situation where the cutoff $M$ is removed as
  $M\rightarrow\infty$.

The equations of motion for the system are given by
\begin{eqnarray}
\partial_t\phi(x,t)&=& \pi(x,t)\nonumber\\
\partial_t\pi(x,t)&=& (\partial_x^2 - 1)\phi(x,t)-  g(\phi(x,t))- r(t)  \alpha(x)
\nonumber\\
dr(t)&=&  -\left(r(t)-\langle\alpha,\pi(t)\rangle\right)dt +
\sqrt{T}d\omega(t).  \label{theeqnonlinear}
\end{eqnarray}
Here $(\phi,\pi)(x,t)$ is a pair of scalar fields; we assume, for
 convenience, that they satisfy periodic
boundary conditions for $x\in[0,2\pi]$.  The fixed vector-valued
function (or distribution) $\alpha=(\alpha_1(x),\alpha_2(x))$ has two components, each in
a Sobolev space $H_s$, that will be further specified below.  The (dependent)
random vector $r(t)= (r_1(t),r_2(t))$ takes values in  ${\mathbb R}^2$; its components are
artifacts of the reservoirs.  The $\alpha_{i}$ couple the oscillators
to the reservoirs, and here $r(t)\alpha\equiv \sum_{i=1}^2 r_i(t)\alpha_i(x)$. The
quantity $\langle \alpha_i,\pi\rangle$ is simply the
$L^2([0,2\pi],\,dx)$-inner product of $\alpha_i(x)$ and
$\pi(x,t)$. The driving term $\omega(t)= (\omega_1,\omega_2)(t) $ is
standard two-dimensional Brownian motion and the $T_i$, $i=1,2$, are the reservoir 
temperatures. The non-linearity $g$ is assumed to be an odd, bounded,
Lipschitz-continuous function. 

The overarching goal of the study of these equations is to construct a stationary (non-equilibrium) process
$\Phi(t)= (\phi(t), \pi(t), r(t))$ governed by the above equations and having physically reasonable sample 
path properties.  In particular, the field component $\phi= \phi(x,t)$ should be
at least square integrable in $x$ a.s., or
actually continuous in $x$ a.s. as in the equilibrium case $T_1= T_2$
and  in the non-equilibrium but linear $g=0$ case, all of which we review below.   

In this article, we establish the existence and uniqueness of
stationary states for these equations with ultraviolet cutoff $M$,
i.e. such that Fourier modes of the field $\phi$ and its momentum
$\pi$ are set to zero for mode numbers $|n|> M$.  We also obtain estimates
on the field covariance uniform in the cutoff with respect to the
stationary cutoff measures. For certain choices 
of the coupling functions $\alpha$, we show 
that $\phi$ is indeed $L^2$-integrable a.s., and that $\|\phi\|_2$ has variance uniformly bounded  in $M$.
  We believe that these estimates are an
important step in establishing bounds on the mode
variances uniform in {\it both} the mode number $n$ and the cutoff
$M$, as one might expect on physical grounds, an issue that we will
address in subsequent work. We conclude with a discussion of
the limit of the cutoff measures, $M\rightarrow \infty$.

It is known that in equilibrium, temperatures $T_1=T_2$, and even for
a non-focusing unbounded nonlinearity $g(x)\propto x^3$, there
is an invariant measure for the system (essentially a Gibbs
state) which is supported on field configurations $\{\phi\}$ that are
almost surely continuous in $x$ (they are  Brownian
motion-like, in fact, H\"{o}lder continuous with any index $<1/2$). The measure is absolutely
continuous with respect to the $g=0$ Gaussian measure (see
\cite{RT4}).  For the non-equilibrium case
$T_1\neq T_2$ but with $g=0$, there is steady-state energy flow, i.e.,there is a random
 variable measuring heat flow
through any fixed point $x\in [0,2\pi]$, and with respect to this
measure, the expectation of this heat flow is non-zero.  Moreover, the
measure has support properties similar to those of the equilibrium
case (see \cite{TW}).  But for the non-linear $g\neq 0$
  non-equilibrium case, the regularity of the field is at
  present not known. If the above equations are to serve as a physical
  model for a non-linear vibrating string in a non-equilibrium
  stationary state, we expect the measure to have support properties on
  field configurations similar to those in the equilibrium case.  In particular,
  we expect the field configurations of $\phi$ to be
  a.s. continuous, not shattered.  It is really this latter problem--
  establishing the regularity of the field with respect to the
  invariant measure-- that is our primary concern  in this article
  and its sequel. The first step in this program is to obtain
  estimates on the field covariance uniform in the ultraviolet cutoff.

We thus consider ultraviolet cutoff versions of the system 
(\ref{theeqnonlinear}) where we retain the Fourier modes $\{\hat{\phi}(n)\}$ for
the field $\phi$ and $\{\hat{\pi}(n)\}$ for momentum $\pi$ for $|n|\leq M$ with $M$ a positive
integer.   Let $P_{\leq M}$ be projection
onto the Fourier modes $\{n\}$, with $|n|\leq M$.  Then setting 
$\Phi_{M}= (\phi_{M},\pi_{M},r_{M})^{T}$
as the cutoff field having non-zero modes only for $|n|\leq M$, $\alpha_{M,i}= P_{M}\alpha_{i}$, we
have that our equations can be written in matrix form as
 \begin{equation}\label{dPhiM.eq}
           d\Phi_{M}= \left(\mAM \Phi_{M} -\mG_{M}(\Phi_{M})\right) dt +\sqrt{\mT}d\omega,        
           \end{equation}
where 
\begin{equation}\label{AMdef.eq}
         \mAM\equiv \left(\begin{array}{cccc}
               0 & 1 & 0 & 0 \\
                \partial_{x}^{2}-1 & 0 & -\alpha_{M,1}(x) & -\alpha_{M,2}(x) \\
                0 &  \langle\alpha_{M,1}& -1 & 0 \\
                0 &  \langle\alpha_{M,2}& 0  & -1 \\
              \end{array}\right),
\end{equation}
with $\langle\alpha_{M,i}$ the linear functional of integration against $\alpha_{M,i}$, and with
\begin{equation}
           -\mG_{M}(\Phi_{M})\equiv \left(\begin{array}{c}
                                     0 \\
                                    -  P_{\leq M}g(\phi_M(x,t)) \\
                                     0 \\
                                     0 \\
                                   \end{array}\right)\,\,{\rm and} \,\,\sqrt{\mT}d\omega\equiv
                                                                 \left(\begin{array}{c}
                                             0 \\
                                             0 \\
                                             \sqrt{T_1}d\omega_1(t) \\
                                             \sqrt{T_2}d\omega_2(t) \\
                                           \end{array}\right).
                                        \label{theeqcutoffm.eq}
                                         \end{equation}
We make the following assumptions about the coupling functions
and the nonlinearity $g$ for the system of equations (\ref{theeqnonlinear}) and 
(\ref{dPhiM.eq}):

\begin{assumption}\label{alphaAssumption} The coupling functions $\alpha_i$, $i=1,2$
are real, with all of their Fourier coefficients $\{\hat{\alpha}_i(n)\}$ 
 non-zero.  There exists a 
positive constant $c_1$ (independent of $n$) such that
\begin{equation}\label{aassume1.eq}
      c_1(|\hat{\alpha}_1(n)|^2+|\hat{\alpha}_2(n)|^2)\leq
      |\hat{\alpha}_1(n)^2+\hat{\alpha}_2(n)^2|,
\end{equation}  
         and there exist both a positive constant $c_2$ and $\theta$,
$-1/2< \theta < 1/2$,
such that, for all $n$,
\begin{equation}\label{aassume2.eq}
   c_2\,(n^2+1)^{\theta/2}\leq |\hat{\alpha}_i(n)|\leq \frac{1}{c_2}\,(n^2+1)^{\theta/2}.
\end{equation}
 The non-linearity $g$ is assumed to be bounded and uniformly Lipshitz.
\end{assumption}

The assumptions on the coupling functions are made in order to assure that
the perturbed eigenvalues of $\mA$ and $\mA_M$ are non-degenerate
and have negative real parts.  In our model, the dissipation of
large $n$ modes is very weak, with rate $\sim-n^{2\theta-2}$.  It is
thus important that the coupling functions themselves have Fourier
coefficients large enough so that the dissipation is not overwhelmed
by the noise driving these modes, and at the same time not so large
that the equations of motion make no sense as the ultraviolet cutoff
is removed.  The assumptions on the $\alpha$'s above appear to capture
the right growth/decay rates in the $\hat{\alpha}(n)$'s.

 The problem of constructing an invariant measure for this {\it
gently} stochastic wave equation may be contrasted with the problem
of constructing stationary states for stochastic parabolic and even
$2$-dimensional stochastic Navier-Stokes equations with viscosity,
where the dissipation of large $n$ Fourier modes of the field is
strong (for Navier-Stokes, see \cite{BKL,EMS,FM,HM,KS}).  For a heat
equation the $n^{th}$ mode decays at a rate $\sim -n^2$, compared
with our rate $\sim -n^{2\theta-2}$ mentioned above.   In
\cite{BdaP}, an invariant measure is obtained for a non-linear wave
equation with cylindrical Brownian motion driving terms but with
strong dissipation for all Fourier modes.  In
\cite{RT4}, an invariant measure is constructed for a one-dimensional
stochastic non-linear Klein-Gordon field but at thermal equilibrium,
with weak dissipation at high Fourier modes; see also \cite{TW} where
an invariant measure for the non-equilibrium but linear case is
given as an invariant measure for an Ornstein-Uhlenbeck process.
See also \cite{LLR}, for an invariant measure for a harmonic crystal
in non-equilibrium.  For invariant measures for wave equations in equilibrium
but with no stochastic driving terms, i.e., equilibrium statistical
mechanics for wave equations, see for example \cite{Mc,Zh}, and for
the non-linear Schr\"{o}dinger equation, see \cite{Bo2,BS,L}.

In the following section, we show existence and uniqueness of invariant measures for
the ultraviolet cutoff systems, and in section 3 we obtain bounds
on the field covariance modes with respect to these invariant measures.
In section 4, we give a general discussion of tightness for these
cutoff measures.
Much of our analysis will use perturbation theory for linear
operators, with estimates on the tails of the
relevant perturbation series. The appendix summarizes these
perturbation calculations.

\section{Existence and Uniqueness of the Stationary Measures for Systems with Ultraviolet Cutoff}

In the following, let ${\mathcal H}$ be the Hilbert space of complex-valued functions $\{\Phi=(\phi,\pi,r)^T\}$
equipped with inner product
\begin{equation}\label{calHdef.eq}
   \langle \Phi_1,\Phi_2\rangle_{\mathcal H}=
   \int_o^{2\pi}(\partial_x\phi_1\partial_x\phi_2^*+\phi_1\phi_2^*
+\pi_1\pi_2^*)\,dx
   +r_1\cdot r_2^*,
\end{equation}
and let $\|\cdot\|_{\mathcal H}$ denote the corresponding norm for ${\mathcal H}$.

The solution $\Phi_M(t)= \Phi_M(t;\Psi)$ to Equation (\ref{dPhiM.eq}) with initial data  $\Phi_M(0)=\Psi$ satisfies the Duhamel integral equation 
\begin{eqnarray}\label{integraleqM.eq}
\Phi_M(t)&=&-\int_0^t e^{(t-s)\mA_M}\mG_M(\Phi_M(s))ds+\int_0^t
                                 e^{(t-s)\mA_M}\sqrt{\mT}d{\bf{\omega}}(s)\nonumber\\&+&e^{t\mA_M}\Psi,\label{mildso}
\end{eqnarray}
(see \cite{SV}, chapter 5).
The field $\Phi_M(t)$ is a vector in the $(4M+4)$-dimensional subspace ${\mathcal H}_M$ of
${\mathcal H}$ spanned by $(2M+1)$ Fourier modes each for $\phi$ and  $\pi$, and by
the two $r$ modes. Using $\Phi_M(t)$, 
one defines a probability semigroup $P_M^t$ acting on bounded
Borel functions on ${\mathbb R}^{4M+4}$ via
\begin{equation}
P_M^tf(\Psi)=E[f(\Phi_{M}(t;\Psi))]
\end{equation}
with $E[\cdot]$ the Brownian motion expectation. The operator $P_M^t$ is a Feller 
semigroup, i.e., $P_M^tf(\Psi)$ is continuous in $\Psi$ if $f$ is
continuous which follows from the integral equation above for
$\Phi_M(t)$ and the assumption that $g$ is Lipschitz.

\begin{proposition}
There exists an invariant measure for the semigroup $P_M^t$.\label{existence}
\end{proposition}
\noindent{\em Proof}: To show the
existence of an invariant measure it is
sufficient to prove that the family of measures $\{\mu_{M,T}\}$ defined by
\begin{equation}
\left\{\int f d\mu_{{M,T}}\equiv\frac{1}{T}\int^T_0P_M^tf(\Psi)dt\right\}_{T>0}
\end{equation}
 is tight on the space ${\mathcal H}_M$ \cite{daPZ}, for $f$'s bounded and continuous.  Since 
 our system is finite dimensional,
one need only show that for each $\Psi\in {\mathcal H}_M$, and $\epsilon>0$, there exist
 an $R>0$ such that for time $T$ sufficiently large,
\begin{equation}
\frac{1}{T}\int^T_0 P\left(\|\Phi_M(t;\Psi)\|_{{\mathcal H}}\geq R\right)dt<\epsilon.
\end{equation}
By the Markov inequality, we have
 \begin{equation}
P\left(\|\Phi_M(t;\Psi)\|_{{\mathcal H}}\geq R\right)\leq {\frac{1}{R}}E
\left[\,\|\Phi_M(t;\Psi)\|_{{\mathcal H}}\right],
\end{equation}
so it suffices to show that $E \left[\|\Phi_M(t,\Psi)\|_{{\mathcal
H}}\right]$ is bounded uniformly in $t$. But from (\ref{mildso}), 
 \begin{eqnarray}
E \left[\|\Phi_M(t;\Psi)\|_{{\mathcal H}}\right]&\leq&
E\left[\left\|\int_0^t
e^{(t-s)\mA_M}(\mG_M(\Phi_M(s)))ds\right\|_{{\mathcal
H}}\right]\nonumber\\&+&E\left[\left\| \int_0^t
e^{(t-s)\mA_M}\sqrt{\mT}d{\bf{\omega}}(s)\right\|_{{\mathcal
H}}\right]\nonumber\\&+&\|e^{t\mA_M}\Psi\|_{{\mathcal
H}}\label{sum1}.
\end{eqnarray}

Now the unperturbed matrix operator $\mB_M$ obtained from $\mA_M$ by setting the  coupling
$\alpha$'s to zero has spectrum
consisting of a doubly degenerate eigenvalue $\lambda_{-1}(0)= -1$,
simple eigenvalues $\lambda_{0}^{\pm}(0)= \pm i$, and  doubly degenerate
eigenvalues $\lambda_n^{\pm}(0)= \pm i\sqrt{n^2+1}$, $1\leq n\leq M$. As seen
in
perturbation theory, the matrix $\mA_M$ has eigenvalues near those
of $\mB_M$, but with small negative real parts (see the Appendix, 
Lemma (\ref{appendixlemma}), Eq.(\ref{eigenshift5appendix.eqn10})).

Given these eigenvalue estimates, it is then easy to see that
the first and third terms on the right side of (\ref{sum1}) are bounded
($\mG_M$ is bounded) uniformly in $t$.  The second term involving the Ito integral
is estimated via Ito calculus, 
\begin{eqnarray}
&&E\left[\left\| \int_0^t e^{(t-s)\mA_M}\sqrt{\mT}d{\bf{\omega}}(s)\right\|_{{\mathcal H}}\right]^2\leq E\left[\left\| \int_0^t e^{(t-s)\mA_M}\sqrt{\mT}d{\bf{\omega}}(s)\right\|^2_{{\mathcal H}}\right]\nonumber\\
&=& \text{\bf Tr}_{{\mathcal H}}\int_0^t e^{s\mA_M}\mT e^{s\mA_M^{\dagger}}ds,
\end{eqnarray}
which is uniformly bounded in $t$, again by the eigenvalue estimates for
$\mA_M$.\qed

\noindent {\it Remark}: The above argument can also
be used to show boundedness of higher moments of the
field $\Phi_{M}$ with respect to the invariant measure.

We next proceed to showing uniqueness of the invariant measure $\mu_M$ satisfying $\mu_MP_M^t= \mu_M$ for the semigroup $P_M^t$.
\begin{proposition} 
Under the assumptions (\ref{alphaAssumption}) on the coupling functions $\alpha$
and nonlinearity $g$, the invariant measure $\mu_M$ for $P_M^t$ is unique.\label{uniqueness}
\end{proposition}
\noindent{\em Proof}: In order to show uniqueness of the invariant
measure, it suffices to show that the system in which the random
field $\Phi_M(t)$ is replaced by a deterministic function $X_M(t)\in
{\mathcal H}_{M}$ satisfying
\begin{equation}\label{controlproblem}
dX_M(t)=(\mA_M X_M(t)-\mG_M(X_M(t)))dt+\sqrt{\bf T}u(t) dt
\end{equation}
is controllable.  This means that, given $X_0$, $X_1$, and a time
$t_1$, one can find a smooth control $u(t) = (u_1(t), u_2(t))$ such
that $X_M(t)$ is a solution to this differential equation, and
$X_M(0)=X_0$, $X_M(t_1)=X_1$ (see \cite{daPZ}, page 144). But
controllability of this system is equivalent to
controllability of the simpler system
\begin{equation}\label{controlproblem2}
dX_M(t)=(\tilde{\mA}_M X_M(t)-\mG_M(X_M(t)))dt+\sqrt{\mT}u(t)dt,
\end{equation}
where the $dr_M(t)$- equations are replaced by $dr_M=\sqrt{\mT}u(t)dt$, i.e.,
 the new and the old controls are related by
\begin{equation}
\sqrt{\mT}u'=-\left(r_M
(t)-\beta\langle\alpha_M,\pi_M(t)\rangle\right)+\sqrt{\mT}u(t),
\end{equation}
and the modified matrix $\tmA_M$ is obtained from $\mA_M$
by replacing its bottom two rows by zero rows.
The solution of equation (\ref{controlproblem2}) satisfies
\begin{equation}
X_M(t)=-\int_0^t\!\!e^{(t-s)\tmAM}\mG_M(X_M(s))ds+\int_0^t\!\!\!e^{(t-s)\tmAM}\smT u(s)ds+ e^{t\tmAM}X_0.\label{solutioncontrol}
\end{equation}

We will show the controllability by first replacing $X_M(s)$ in the
first integral on the right side of this last equation
(\ref{solutioncontrol}) by an arbitrary continuous function $Z_M(s)$,
thereby obtaining an expression for $u(t)$ explicitly in terms of $Z_M(s)$, and
then by using a fixed point theorem. We thus consider
\begin{equation}\label{solutioncontrolr}
X_M(t)=-\int_0^t\!\!e^{(t-s)\tmAM}\mG_M(Z_M(s))ds
+\int_0^t\!\!\!e^{(t-s)\tmAM}\smT u(s)ds+ e^{t\tmAM}X_0,
\end{equation}
for which we seek a control $u(t)$ such that for fixed $Z_M(s)$, $X_M(t_1)=X_1$.\par
Let
\begin{equation}\label{cont.eq}
\mD_t=\int_0^te^{-s\tmAM}\mT e^{-s\tmAM^{\dagger}}ds.
\end{equation}
\begin{lemma}\label{Dinvert.lemma}
The matrix $\mD_{t}$ is a non-negative self-adjoint invertible $(4M+4)\times(4M+4)$ matrix for
$t>0$, provided the Fourier coefficients of $\alpha_1$ and $\alpha_2$
satisfy the assumption Eq.(\ref{aassume1.eq}).
\end{lemma}

\noindent{\em Proof}: Clearly, the integrand of the integral in (\ref{cont.eq}) is non-negative and
continuous in $s$.  If $D_t$ were not invertible, then for some $X\neq 0$ , 
$\langle X, \mD_{t}X\rangle_{\mathcal H}= 0$, and in particular
$\smT e^{-s\tmAM^{\dagger}}X=0$ for all $0\leq s \leq t$. Since
$\smT e^{-s\tmAM^{\dagger}}X$ is smooth in $s$, we have
${\frac{d^n}{ds^n}}\left(\smT e^{-s\tmAM^{\dagger}}X\right){\Big
|}_{s=0}=0$ for all $n$, so 
$\smT\tmAM^{\dagger\,\,n}X=0,\ \ n=0,1,2,\cdots$.
However we know that for any $Y\in {\mathcal H}_M$, $\langle Y,\smT(\tmAM^{\dagger})^nX\rangle_{\mathcal H}=0$,
 which implies that
\begin{equation}
 \langle(\tmAM)^n\smT Y,X\rangle_{\mathcal H}=0\end{equation}
 for all $Y$, and thus the range of
 $\{(\tmAM)^n\smT\}_{n=0}^{\infty}$ is not the whole space ${\mathcal H}_M$. But this contradicts the following claim.

\begin{claim}
Vectors consisting of the columns of
$\{(\tmAM)^n\smT\}_{n=0}^{\infty}$ $n=0,1,2,\dots$ span  ${\mathcal H}_M$.\label{Homander}
\end{claim}

\noindent {\it Remark}: The generator of the Ornstein-Uhlenbeck semigroup defined
 by the linear equation $d\Phi_M(t)={\tmAM} \Phi_M(t)dt+\sqrt{\mT}d\omega$
is 
\begin{eqnarray}
\mathcal{G}&=&T_1{\frac{\partial^2}{\partial r_{1}^2}}+T_2\frac{\partial^2}{\partial r_{2}^2}
+\int dx \,\pi_M\frac{\delta}{\delta \phi_M}\nonumber\\&+&
\int dx \left((\partial_x^2-1)\phi_M+\beta r\alpha_M\right)
\frac{\delta}{\delta \pi_M}\,.
\end{eqnarray}
If we denote 
\begin{eqnarray}
\mathbf{X}_1&=&\frac{\partial}{\partial r_{1}},\,\,\,
\mathbf{X}_2=\frac{\partial}{\partial r_{2}},\,\, {\rm and}\nonumber\\
\mathbf{Y}&=&\int dx \left(\pi_M\frac{\delta}{\delta\phi_M}+
 \left((\partial_x^2-1)\phi_M+\beta r \alpha_M\right)
\frac{\delta}{\delta\pi_M}\right),\end{eqnarray}
then the conclusion in claim
(\ref{Homander}) is equivalent to the condition that the Lie algebra
generated by the vector fields $\mathbf{X}_1$, $\mathbf{X}_2$, and
$\mathbf{Y}$ has full rank at each point of ${\mathbb R}^{4M+4}$, and this
implies that $\mathcal{G}$ is hypoelliptic(See \cite{Ho}).\\

\noindent{\em Proof of Claim}: Proving that the columns of
$\{\tmAM^n\smT\}_{n=0}^{\infty}$ span ${\mathcal H}_M$ amounts to
showing, in the end, that $\{P_{\leq
M}(\partial_x^2-1)^k\alpha_1\}_{k=0}^{\infty}$ and $\{P_{\leq
M}(\partial_x^2-1)^k\alpha_2\}_{k=0}^{\infty}$ span $P_{\leq M}L^2[0,2\pi]$. They
do so provided that all Fourier coefficients $\hat{\alpha}_{1}(m)$ and $\hat{\alpha}_{2}(m)$ of 
the $\alpha$'s are nonzero and that $(\hat\alpha_1(m),
\hat\alpha_1(-m))$ and $(\hat\alpha_2(m), \hat\alpha_2(-m))$ are linearly
independent, $m\neq 0$.  But this is implied by the assumption Eq.(\ref{aassume1.eq}). This concludes the proof
of the claim and hence the Lemma (\ref{Dinvert.lemma}). \qed

Returning now to the proof of Proposition (\ref{uniqueness}), we identify
$u(t)$ with the column vector having the four components $(0,0,u_{1}(t),u_{2}(t))$,
and we set
\begin{eqnarray}\label{ucontrol.eq}
  u(t)=&-&\smT e^{-t\tmAM^{\dagger}}
\mD_{t_1}^{-1}\times\nonumber\\
&&\left(X_0-\int_0^{t_1}e^{-s\tmAM}\mG_M(Z_M(s))ds
-e^{-t_1\tmAM}X_1\right).\end{eqnarray}
Putting $u(t)$ into equation (\ref{solutioncontrolr}), we have
\begin{eqnarray}
X_M(t)&=&-\int_0^te^{(t-s)\tmAM}\mG_M(Z_M(s))ds\nonumber\\
&&-e^{t\tmAM}\mD_{t}\mD_{t_{1}}^{-1}\left(X_0-\mu\int_0^{t_1}e^{-s'\tmAM}\mG_M(Z_M(s'))ds'
-e^{-t_1\tmAM}X_1\right)\nonumber\\
&&+e^{t\tmAM}X_0.\label{solutioncontrolu}
\end{eqnarray}
It is then easy to see from this equation that $X_M(t_1)=X_1$.

We consider the right side of this last equation
(\ref{solutioncontrolu}) as a nonlinear operator $\mK(Z_M(\cdot))$
which maps the Banach space ${\bf C}_{[0,t_1]}({\mathcal H}_M)$
(continuous functions mapping $[0,t_1]$ to ${\mathcal H}_M$) into itself,
with $X_M(t)=\mK(Z_M)(t)$. Let $\Theta=\{Z\in {\bf C}_{[0,t_1]}({\mathcal
H}_M):\,\|Z(t)\|_{\infty} \leq C\}$ with $C$ a uniform bound of the
right hand side of (\ref{solutioncontrolu}).  Then $\Theta$ is closed
and convex. Denote the range of the operator $\mK$ by $\Omega$; then
$\Omega\subset\Theta$, and functions in $\Omega$ are equicontinuous in
$t$ ($\mG_M$ is bounded), so $\Omega$ is compact by the Arzela-Ascoli
theorem. By Schauder's fixed point theorem, (\cite{RS}, page 151),
$\mK$ has a fixed point $X^{\ast}_M(t)$ such that
$X^{\ast}_M(t)=\mK(X^{\ast}_M)(t)$, which satisfies the conditions
$X^{\ast}_{M}(0)= X_{0}$ and $X^{\ast}_{M}(t_{1})= X_{1}$.  Here
$X^{\ast}_M$ is a solution of Eq.(\ref{solutioncontrolr}) with control $u(t)$ given
by Eq.(\ref{ucontrol.eq}).  This concludes the proof of the
controllability and hence the uniqueness of the invariant measure, Proposition (\ref{uniqueness}).\qed

\section{Estimates on the covariance of the field in the stationary state}

Let $\Phi_M(x,t)=(\phi_M(x,t),\pi_M(x,t),r_{1,M}(t),r_{2,M}(t))^T$ be the stationary field
 corresponding to the M-cutoff stochastic differential equation (\ref{dPhiM.eq}) with invariant
measure $\mu_M$. We write $f^{\pm}_{M,n,\sigma}=(f^{\pm}_{M,n,\sigma,\phi},\,f^{\pm}_{M,n,\sigma,\pi},\,f^{\pm}_{M,n,\sigma,r})$ as a left eigenvector
for $\mA_M$, $f^{\pm}_{M,n,\sigma}\mA_M= \lambda_{M,n,\sigma}^{\pm}$, 
expressing $f^{\pm}_{M,n,\sigma}$ in its components, and similarly we write
$e^{\pm}_{M,n,\sigma}=(f^{\pm}_{M,n,\sigma,\phi},\,f^{\pm}_{M,n,\sigma,\pi},\,f^{\pm}_{M,n,\sigma,r})^T$
for the column components of the right eigenvector corresponding to
the same eigenvalue $\lambda^{\pm}_{M,n,\sigma}$. The eigenvectors
are normalized so that 
\begin{eqnarray}\label{innerprod.eq}
\lefteqn{\langle f^{\pm}_{M,n,\sigma},e^{\pm}_{M,n,\sigma}\rangle}&&\\
&\equiv& \int\,dx\left( f^{\pm}_{M,n,\sigma,\phi}(x)e^{\pm}_{M,n,\sigma,\phi}(x)+f^{\pm}_{M,n,\sigma,\pi}(x)e^{\pm}_{M,n,\sigma,\pi}(x)\right) +f^{\pm}_{M,n,\sigma,r}e^{\pm}_{M,n,\sigma,r} =1\nonumber
\end{eqnarray}
 (making $e^{\pm}_{M,n,\sigma,\phi}\otimes f^{\pm}_{M, n,\sigma,\phi}$ the kernel of 
a projection), and, for the sake of definiteness, the $\pi$-component 
$f^{\pm}_{M, n,\sigma,\pi}$ has $L^2[0,2\pi]$-norm $1/\sqrt{2}$. For this
 normalization, asymptotically for $n$ and $M$ large, $e^{\pm}_{M, n,\sigma,\pi}$
 has the same $1/\sqrt{2}\,\,\,$ normalization.

In this section, we obtain estimates on the covariance of $\Phi_M(t)$
at equal times with respect to the invariant measure $\mu_M$, 
$E_{\mu_M}[\Phi_M(f^{\pm}_{M,n,\sigma},t)\,\Phi_M(f^{\pm}_{M,m,\sigma'},t)^*]$,
where
\begin{eqnarray}       
        \lefteqn{\Phi_M(f,t)= \langle f,\Phi_M(t)\rangle}&&\\
&\equiv& \int\,dx\left( f^{\pm}_{M,n,\sigma,\phi}(x)\phi(x,t)+f^{\pm}_{M,n,\sigma,\pi}(x)\pi(x,t)\right) +f^{\pm}_{M,n,\sigma,r}r(t).  \nonumber
\end{eqnarray}
These covariances are independent of time, and so we henceforth suppress
the time $t$ in their expressions.

 The variances of the $r_M$ variables are bounded by the average of 
the temperatures, as implied by the following identity.  
\begin{lemma}\label{rvar.lemma}
 We have 
 \begin{equation}\label{rvar.eq}
E_{\mu_M}[r_M^2] \equiv E_{\mu_M}[r_{1,M}^2+r_{2,M}^2]=\frac{1}{2}(T_1+T_2).
\end{equation}
\end{lemma}
\noindent {\it Remark}: The lemma says that the average expected energy
of each of the $r$ variables is given by one-half the average of the temperatures, as one
 might anticipate from equipartition of energy in equilibrium.\\  

\noindent {\em Proof}: Let $H_M(t)$ be the (degenerate) Liapunov function defined
\begin{equation}
   H_M(t)= \frac{1}{2}\|\Phi_M(t)\|^2_{\mathcal H}+\int_0^{2\pi}d\,x\, h(\phi_M(x,t))
\end{equation}
with $h(x)$
an antiderivative of $g(x)$.  The Ito differential of this quantity is given by
  \begin{equation}
    dH_M(t)= \left(\frac{1}{2}(T_1+T_2)-r_M^2\right)dt +r_M\sqrt{T} d\omega(t). 
   \end{equation}
By stationarity of $\Phi_M$, we have that $dE_{\mu_M}[H(t)]=0$, and so 
by the non-anticipating property of $r_M(t)$ and thus independence
of $r_M(t)$ and $d\omega$, 
\begin{equation}
    0 = E_{\mu_M}[d H_M(t)]= E_{\mu_M}[\frac{1}{2}(T_1+T_2)-r_M^2]\,dt= 0.
\end{equation}
\qed

The next task is to obtain an estimate on the covariance of different modes of
the field. In the
following, $\|f^{\pm}_{M,n,\sigma,\pi}\|_2$ is the $L^2[0,2\pi]$-norm of
$f^{\pm}_{M,n,\sigma,\pi}$,  and $\langle f^{\pm}_{M,n,\sigma,\pi},e\rangle$ is the
$L^2[0,2\pi]$-inner product of $f^{\pm}_{M,n,\sigma,\pi}$ with any other  $L^2[0,2\pi]$-function $e(x)$.
\begin{lemma}\label{etcov.eq} (Equal time covariance).
 We have, for $n\neq m$, or for $n=m$ and the $\pm$'s differing, 
 \begin{eqnarray}\label{eqt1.ineq}
\lefteqn{\left|E_{\mu_M}[\Phi_M(f^{\pm}_{M,n,\sigma})\Phi_M(f^{\pm'}_{M,m,\sigma'})^*]\right|}&&\\
&\leq& \frac{1}{|\lambda^{\pm}_{M,n,\sigma}+\lambda^{\pm'\,\,*}_{M,m,\sigma'}|}
\left(E[|\Phi_M(f^{\pm}_{M,n,\sigma})|^2]^{1/2}\|f^{\pm'}_{M,m,\sigma',\pi}\|_2 
\|g\|_{\infty}\phantom{\left|\langle f^{\pm}_{M,n,\sigma},T f^{\pm'\,\,\dagger}_{M,m,\sigma'}\rangle_{\mathcal H}\right|}\right.\nonumber\\
&& \left. +E[|\Phi_M(f^{\pm'}_{M,m,\sigma'})|^2]^{1/2}\|f^{\pm}_{M,n,\sigma,\pi}\|_2 \|g\|_{\infty}
 + \left|\langle f^{\pm}_{M,n,\sigma},T f^{\pm'\,\,\dagger}_{M,m,\sigma'}\rangle_{\mathcal H}\right|\right).\nonumber
\end{eqnarray}
\end{lemma}

\noindent {\em Proof}: Again by stationarity and by using Eq.(\ref{dPhiM.eq}), we have that
\begin{eqnarray}
  \lefteqn{d E_{\mu_M}[\Phi(f^{\pm}_{M,n,\sigma})\Phi(f^{\pm'}_{M,m,\sigma'})^*]= 0}&&\nonumber\\
  &=& (\lambda^{\pm}_{M,n,\sigma}+\lambda^{\pm'\,\,*}_{M,m,\sigma'})E_{\mu_M}[\Phi_M(f^{\pm}_{M,n,\sigma})\Phi_M(f^{\pm'}_{M,m,\sigma'})^*]\,dt \nonumber\\
&& -\left(E_{\mu_M}[\Phi_M(f^{\pm}_{M,n,\sigma}) \langle f^{\pm'}_{M,m,\sigma',\pi}, g\rangle^*]
   +E_{\mu_M}[\langle f^{\pm}_{M,n,\sigma,\pi}, g\rangle \Phi_M(f^{\pm'}_{M,m,\sigma'})^*]\right.\nonumber\\
&&\phantom{XXXX} \left.  + \langle f^{\pm}_{M,n,\sigma},T f^{\pm'\,\,\dagger}_{M,m,\sigma'}
\rangle_{\mathcal H}\right) dt.      
\end{eqnarray}
Ineq.(\ref{eqt1.ineq}) follows immediately.  \qed

Finally, we estimate the variances of individual modes (actually their sum). We use 
the notation
$ |e^{\pm}_{M,n,\sigma,r}|= \left(e^{\pm}_{M,n,\sigma,r}\cdot e^{\pm\,*}_{M,n,\sigma,r}\right)^{1/2},
$
where we have made explicit the dot product of the two $r_M$ components 
for $e^{\pm}_{M,n,\sigma,r}$.

\begin{proposition}\label{rrbound.prop}
   There exists a constant $C>0$ independent of the
ultraviolet cutoff $M$ such that
 \begin{equation}\label{rrbound3.ineq}
       \sum_{n,\sigma,\pm}|e^{\pm}_{M,n,\sigma,r}|^2\,E\left[|\Phi_M(f^{\pm}_{M,n,\sigma})|^2\right]
\leq  C.
\end{equation}
The weight $|e^{\pm}_{M,n,\sigma,r}|^2$ is comparable
to $(n^2+1)^{\theta-1}$; there exists a constant $c_1$ with
\begin{equation}\label{weight.eq}
          c_1(n^2+1)^{\theta-1} \leq   |e^{\pm}_{M,n,\sigma,r}|^2\leq c_1^{-1}(n^2+1)^{\theta-1}.
\end{equation}
\end{proposition}

\noindent {\it Proof}: We begin by writing $r_M= (r_{M,1},r_{M,2})^T$ in an eigenfunction expansion, 
   \begin{equation}\label{rspectral.eq}
     r_M= \sum_{n,\sigma,\pm}e^{\pm}_{M,n,\sigma,r}\Phi_M(f^{\pm}_{M,n,\sigma}),
\end{equation}
(the expansion is complete!) and denoting the sum
on the left side of the  proposition inequality (\ref{rrbound3.ineq})
 by $\| \hat{r}_M\|_{\ell^2}^2$.   
By Lemma (\ref{rvar.lemma}) above and
the expansion for $r_M$, we have that
\begin{eqnarray}\label{doublesum.eq}
 \lefteqn{\frac{1}{2}(T_1+T_2) = E[r_M^2]}&&\\
&=& \| \hat{r}_M\|_{\ell^2}^2
\nonumber\\
&&+ \sum_{(n,\sigma,\pm)\,\,\neq}\sum_{(m,\sigma',\pm')}\left(e^{\pm\,\,*}_{M,n,\sigma,r}\cdot e^{\pm'\,\,*}_{M,m,\sigma',r}\right)\,E\left[\Phi_M(f^{\pm\,\,*}_{M,n,\sigma})\Phi_M(f^{\pm'\,\,*}_{M,m,\sigma'})\right],\nonumber
\end{eqnarray}  
where the  double sum is over off-diagonal terms,
$n\neq m$, or the $\pm$'s different, or $\sigma\neq \sigma'$.

For this equation, we use Lemma (\ref{etcov.eq}) to estimate the  double sum over the non-resonant terms 
with $n\neq m$ or the $\pm$'s differing.  For these terms, the denominators
 $|\lambda^{\pm}_{M,n,\sigma}+\lambda^{\pm'\,\,*}_{M,m,\sigma'}|$ behave like $|n-m|$ if $n\neq m$
 and $\pm= \pm'$, or like $|n+m|$ if $\pm=\mp'$, and so in either case these denominators are not 
 dangerous (see Lemma \ref{appendixlemma} of the appendix).  Thus, the ``kernel'' $|\lambda^{\pm}_{M,n,\sigma}+\lambda^{\pm'\,\,*}_{M,m,\sigma'}|^{-1}$ is $\ell^p$-summable
  for any $p>1$.   We also use the fact that  $e^{\pm}_{M,n,\sigma,r}= \langle\alpha,e^{\pm}_{M,n,\sigma,r}\rangle/(1+\lambda^{\pm}_{M,n,\sigma})
 ={\mathcal O}(n^{\theta-1})$,  by Eqs.(\ref{esigmaapp.eq},\ref{Aae.ineq}) of the appendix, which happens to establish
 Eq.(\ref{weight.eq}) of the Proposition as well. It follows that $|e^{\pm}_{M,n,\sigma,r}|$ is $\ell^p$-summable for
$p>1/(1-\theta)$. These estimates, together with Young's inequality, then show that the non-resonant
part of the double sum in Eq.(\ref{doublesum.eq}) is bounded below by
\begin{equation}\label{doublesum.eq1}
  - c_1 \| \hat{r}_M\|_{\ell^2} \left(\sum_{n,\sigma,\pm}|e^{\pm}_{M,n,\sigma,r}|^p\right)^{1/p}\geq -c\| \hat{r}_M\|_{\ell^2}
\end{equation}
for suitable positive constants $c, c_1$ and suitably large $p<2$.

The near-resonant terms in the double sum of Eq.(\ref{doublesum.eq}),  terms with $n=m$ and $\pm$'s the same  but $\sigma\neq\sigma'$, are
more delicate. Fortuitously $e^{\pm}_{M,n,\sigma,r}$
and $e^{\pm}_{M,m,\sigma',r}$ are nearly orthogonal for $n\rightarrow \infty$, i.e.,
\begin{equation}
        |e^{\pm,\,\,*}_{M,n,\sigma,r}\cdot e^{\pm}_{M,n,\sigma',r}|= |e^{\pm,\,\,*}_{M,n,\sigma,r}|\,|e^{\pm,\,\,*}_{M,n,\sigma',r}|\times {\mathcal O}\left(n^{2\theta-1}\ln{ n}\right),
\end{equation}
by Lemma (\ref{appendixlemma}),  Eq.(\ref{Vorthog.appendix.eq11}) of the appendix.  This inequality implies that for some fixed $N$ (independent of the ultraviolet
cutoff and chosen so that the ${\mathcal O}\left(n^{2\theta-1}\ln{ n}\right)$- factor
is $< 1/2$ for $n\geq N$), the tail series satisfies 
\begin{equation} \label{doublesum.eq2}
    \sum_{\{(n,\sigma,\pm): \,n\geq N\}}\left(e^{\pm}_{M,n,\sigma,r}\cdot e^{\pm\,\,*}_{M,n,\sigma}\right)\,E\left[\Phi(f^{\pm}_{M,n,\sigma})\Phi(f^{\pm\,\,*}_{M,n,\sigma',r})\right] \geq - \frac{1}{2}\|\hat{r}_M\|^2_{\ell^2}.
\end{equation}
Now the sum of the variances of the low $n$ modes $n\leq N$ is bounded by
a constant $c_2$, since the variance of an individual mode $\Phi_M(f^{\pm}_{M,n,\sigma})$
is certainly bounded by $ {\rm const}/ ({\rm Im}\lambda^{\pm}_{M,n,\sigma})^{2}$, as
seen from the Duhamel integral representation, Eq.(\ref{integraleqM.eq}), for $\Phi_M$  in
 the $t\rightarrow \infty$ limit. Thus, combining Eq.(\ref{doublesum.eq}) and Ineqs.(\ref{doublesum.eq1},\ref{doublesum.eq2}),
we obtain a quadratic inequality for $\| \hat{r}_M\|_{\ell^2}$, 
\begin{equation}\label{doublesum.eq3}
 \frac{1}{2}(T_1+T_2) \geq \| \hat{r}_M\|_{\ell^2}^2-c\| \hat{r}_M\|_{\ell^2} 
    - \frac{1}{2}\|\hat{r}_M\|^2_{\ell^2}-c_2,
\end{equation}
which gives the bound of the proposition.  \qed

     In the inequality (\ref{rrbound3.ineq}) of  Proposition
(\ref{rrbound.prop}) we can actually replace function
$f^{\pm}_{M,n,\sigma}$ in the expectation
$E\left[|\Phi(f^{\pm}_{M,n,\sigma})|^2\right]$ with the
free eigenfunctions $f^{\pm}_{M,n,\sigma}(0)$, that is, left
eigenfunctions of $\mB_M$, i.e. the matrix operator  obtained from $\mA_M$ by setting
the
coupling functions $\alpha$ to zero.  For $n\geq 0$, we
take $f^{\pm}_{M,n,\sigma}(0)= \frac{1}{\sqrt{4\pi}}(\pm
i(n^2+1)^{-1/2}e^{i\sigma n x},e^{i\sigma n x}, 0)$, with $\sigma=
1\,\,{\rm or}\,\,-1$ (except when $n=0$, where there is no
$\sigma$-dependence), and $f_{M,-1,\sigma}(0)= (0,0,1,0)$ or
$(0,0,0,1)$.
\begin{corollary}\label{ffbound.cor}
   There exists a finite constant $C>0$ independent of the cutoff $M$
   such that
 \begin{equation}\label{ffbound3.ineq}
       \sum_{n,\sigma,\pm}(n^2+1)^{\theta-1} E\left[|\Phi_M(f^{\pm}_{M,n,\sigma}(0))|^2\right]
\leq  C.
\end{equation}
\end{corollary}

\noindent {\it Remark}: The corollary implies that
   \begin{equation}
       \sum_n (n^2+1)^{\theta}E[|\hat{\phi}_M(n)|^2]\leq C
   \end{equation}
uniformly in $M$, with $\hat{\phi}_M(n)= \int e^{-inx}\phi_M(x)$ the usual Fourier coefficient of the field component $\phi_M$ of $\Phi_M$. If $\theta\geq 0$,
then the field $\phi_M$ is in $L^2$ a.s., with
the variance of $\|\phi_M\|_2$ uniformly bounded in $M$; obviously,
each mode $\hat{\phi}_M(n)$ has variance uniformly bounded in $n$. 
However, we believe that this bound can be improved, that in fact
${\rm var}(\hat{\phi}_M(n))= {\mathcal O}(n^{-2})$ and that ${\rm var}(\hat{\pi}_M(n))=
{\mathcal O}(1)$ uniformly in $M$ as in the linear and
non-linear equilibrium cases
and the  linear $g=0$ non-equilibrium case.  This improvement will be the
subject of a subsequent investigation. \\ 

\noindent {\it  Proof}:  We expand 
\begin{equation}\label{Phiexpansion.eq}
   \Phi_M(f^{\pm}_{M,n,\sigma}(0))= \sum_{m,\sigma',\pm'}\langle f^{\pm}_{M,n,\sigma}(0),
e^{\pm'}_{M,m,\sigma'}\rangle\Phi_M(f^{\pm'}_{M,m,\sigma'}). 
\end{equation}
The coefficients of this expansion (with inner products as in Eq.(\ref{innerprod.eq})) satisfy
   \begin{equation}\label{ineqtable.eq}
     \langle f^{\pm}_{M,n,\sigma}(0),e^{\pm'}_{M,m,\sigma'}\rangle=\left\{
\begin{array}{l} {\mathcal O}(m^{\theta-1}),\,\, n= -1, m\rightarrow \infty\\
   {\mathcal O}(n^{\theta-1}),\,\,  m= -1, n\rightarrow\infty\\
   {\mathcal O}\left(\frac{n^{\theta}m^{\theta-1}}{(n-m)}\right),\,\, m\neq n,\,\,m,n\rightarrow\infty\\
   {\mathcal O}(1),\,\,{\rm otherwise},
\end{array}\right.
\end{equation}
all uniform in the cutoff $M$ (see Eq.(\ref{f0eappendix.eq10}) where
these estimates are shown).  
Substituting the expansion Eq.(\ref{Phiexpansion.eq})
into $E[|\Phi_M(f^{\pm}_{M,n,\sigma}(0))|^2]$, we obtain the double sum
of terms 
\begin{equation}\label{mmterms.eq}
   \langle f^{\pm}_{M,n,\sigma}(0),e^{\pm'\,\,}_{M,m,\sigma'}\rangle\langle f^{\pm}_{M,n,\sigma}(0),e^{\pm''}_{M,m',\sigma''}\rangle^* E[\Phi_M(f^{\pm'}_{M,m,\sigma'})\Phi_M(f^{\pm''}_{M,m',\sigma''})^*].
\end{equation}
Off-diagonal terms $m\neq m'$ must be estimated using the non-resonant Ineq.(\ref{eqt1.ineq}) of
Lemma (\ref{etcov.eq}) and 
the above inequalities (\ref{ineqtable.eq}). For $m,m'\rightarrow\infty$ these 
terms are  
   \begin{eqnarray}
     \lefteqn{{\mathcal O}\left(\frac{n^{2\theta}m^{\theta-1}m'^{\,\theta-1}}{(n-m)(n-m')(m-m')}\right)}&&\\
 &&\times \left(E[|\Phi_M(e^{\pm'}_{M,m,\sigma'})|^2]^{1/2} +E|\Phi_M(f^{\pm''}_{M,m',\sigma''})|^2]^{1/2}+{\mathcal O}(m^{\theta-1}m'^{\theta-1})\right).\nonumber
\end{eqnarray}

   It remains to sum the on- and off-diagonal terms (\ref{mmterms.eq})
over $m,m'$, and over $n$ with the prefactor $n^{2\theta -2}$. One uses the fact that
 $m^{\theta-1}E[|\Phi_M(fe^{\pm'}_{M,m,\sigma'}|^2]^{1/2}$ is
$\ell^2$-summable in $m$, the content of Proposition (\ref{rrbound.prop}). 
The near-resonant $m=m'$ terms are estimated using the inequalities 
(\ref{ineqtable.eq}). 
By extensive use of Young's inequality and H\"{o}lder's inequality, the triple sum
is shown to be finite. \qed  

\section{On the tightness of the ultraviolet cutoff stationary measures} 

Let $\{\mu_M\}$ denote the unique stationary measures for the 
ultraviolet cutoff systems of the previous sections, $M$ labeling the cutoff.
 Let $\Phi_M(t)$ be the canonical stationary process associated with $\mu_M$ so that
in particular $\mu_M$ is the law for $\Phi_M(t)$ for any time $t$. It will
be convenient to regard the field $\Phi_M(t)$ as taking values 
in a space of distributions dual to the Schwartz space ${\mathcal S}\equiv
{C}_{\rm per}^{\infty}[0,2\pi]\times {C}_{\rm
per}^{\infty}[0,2\pi]\times {\mathbb R}\times {\mathbb R}$, but
where the Fourier modes $\{\hat{\phi}_M(n),\,\hat{\pi}_M(n)\}$ of $\Phi_M$ are all zero for $|n|>M$.

 Corollary (\ref{ffbound.cor}) provides a bound on the variance
 of the Fourier mode of the field, 
$var_{\mu_M}(\Phi_M(f_{n,\sigma}^{\pm}(0))\sim n^{2-2\theta}$ uniform in $M$ at a fixed time,
 say $t=0$.  As before,
we suppress the explicit time dependence. Let ${\mathcal H}_s= H_{s}\oplus H_{s-1}\oplus C\oplus C$. 
  Then generally,
 we note the following:  

\begin{proposition}\label{tightness.lem} Assume that the fields $\Phi_M$ are of mean zero with
respect to $\mu_M$ for all ultraviolet cutoffs $M$, and that the field modes are of variance
\begin{equation}\label{varbound.eqn}
     var_{\mu_M}[\Phi_M(f_{n,\sigma}^{\pm}(0))]\equiv E_{\mu_M}[\, |\Phi_M(f_{n,\sigma}^{\pm}(0))|^2]\leq C (1+n^2)^p
\end{equation}
for some constants $C$ and power $p$ independent of $M$. Then the measures
$\{\mu_M\}$ are tight in the weak-$*$ sense that there exist a subsequence
$\{\mu_{M_j}\}$, $j= 1,2,...$, a limiting measure $\mu$, and a limiting field
$\Phi$ 
 such that
        \begin{eqnarray}
           \lefteqn{\lim_{j\rightarrow \infty} E_{\mu_{M_j}}[ F(\Phi_{{M_j}}(f_1),
 \Phi_{{M_j}}(f_2),...
,\Phi_{{M_j}}(f_k))]}\nonumber\\
 &&\phantom{XX}= E_{\mu}[ F(\Phi(f_1), \Phi(f_2),...
,\Phi(f_k))]
\end{eqnarray}
for all bounded continuous functions $F(x_1,x_2,\cdots,x_k)$ on ${\bf
R}^k$, and with $f_1,f_2,...f_k\in {\mathcal S}$ for all $k$. The limiting measure
has support in the space of distributions ${\mathcal S}'$ and is
$\sigma$-additive on the Borel $\sigma$-algebra generated by
cylinder sets of the form
    \begin{equation}
        \{\Phi: (\Phi_M(f_1), \Phi_M(f_2),...
,\Phi_M(f_k))\in B\}
    \end{equation}
with base $f_1,f_2,..f_k\in {\mathcal S}$ and  $B$ a Borel set in ${\mathbb R}^k$.

We have that for $s< -p-1/2$
      \begin{equation}\label{Einftybnd.eqn}
  E_{\mu}[\|\Phi\|_{{\mathcal H}_s}^2]< C_1
      \end{equation}
for some finite constant $C_1$; the same bound holds for the $\mu_M$'s. In
particular, $\Phi$ is in ${\mathcal H}_s$ a.s.
\end{proposition}

\noindent {\it Remark}: Again, if the Fourier coefficients of the $\alpha$'s behave as a
 power, $|\hat\alpha(n)|\sim n^{\theta}$, then the variances of
the Fourier modes indeed satisfy  Ineq.(\ref{varbound.eqn}), and so we obtain a limiting
 measure for which $\Phi\in {\mathcal H}_s$, $s< \theta-3/2$.  However, we do not claim that
this measure is an invariant measure for the nonlinear stochastic wave equation;
 this remains an open problem.\\

\noindent{\it Proof}: The {\it marginals} of $\{\mu_M\}$
restricted to functions $F$ just depending on a fixed and finite number of
Fourier modes $\{\Phi_M(f_{n,\sigma}^{\pm}(0))\}_{\{n:\,|n|\leq n_0\}}$ are tight.  This is the case since a closed ball in a
finite-dimensional Euclidean space is compact, and given
$\epsilon>0$,
    \begin{eqnarray}
       P_{\mu_M}\{\Phi: \|P_{\{|n|\leq n_o\}}\Phi\|_2>R\}&\leq& \frac{1}{ R^2}
E_{\mu_M}[\sum_{|n|\leq n_0,\sigma,\pm}|\Phi_M(f_{n,\sigma}^{\pm}(0))|^2]\nonumber\\ &\leq& \frac{c}{ R^2}\sum_{|n|\leq
n_0,\sigma,\pm}(1+n^2)^p<\epsilon
      \end{eqnarray}
for another constant $c$ for $R$ sufficiently large, by our bounds on
the Fourier coefficient variances (see \cite{Sh}). Here, $P_{\{|n|\leq n_o\}}\Phi$ is projection of $\Phi$ onto its Fourier modes with $|n|\leq n_o$.   By
this tightness, one can construct a subsequence $\{\mu_{M_{j,n_0}}\}$ with convergent
marginals based on the random variables $\{\Phi_{M_j}(f_{n,\sigma}^{\pm}(0))\}_{\{n:\,|n|\leq n_0\}}$.

One then passes to a sub-subsequence to get convergence for
marginals based on a larger collection $\{\Phi_M(f_{n,\sigma}^{\pm}(0))\}$, $|n|\leq n_1$ for $n_1>n_0$ and
diagonalization, one obtains a subsequence $\{\mu_{M_j}\}$ which,
integrated against any continuous function $F$ of a finite number of
Fourier modes, converges, i.e., $\lim_{j\rightarrow\infty} E_{\mu_{M_j}}(F)$ exists.

Again by Chebyshev, given $\epsilon$
\begin{eqnarray}\label{Cheb1:eqn}
       \lefteqn{P_{\mu_M}\{\Phi: \|\Phi\|_{{
\mathcal H}^{-p-1}}>R\}}\nonumber\\
&\leq& \frac{1}{ R^2}
E_{\mu_M}\left[\sum_{n,\sigma,\pm}\frac{1}{(1+n^2)^{p+1}}|\Phi_M(f_{n,\sigma}^{\pm}(0))|^2\right]\nonumber\\
 &\leq& \frac{C}{ R^2}\sum_{n,\sigma,\pm}\frac{(1+n^2)^p}{(1+n^2)^{p+1}}<\epsilon
      \end{eqnarray}
for $R$ sufficiently large, uniformly in $M$.
This implies the last inequality of the lemma, Ineq.(\ref{Einftybnd.eqn}).

It is then easy to see that the domain of definition of $\mu$
extends uniquely to functions of the form $F(\Phi(f_1), \Phi(f_2),...
,\Phi(f_k))$,
with $f_1,f_2,...,f_k\in {\mathcal S}$, $F(x_1,x_{2},...,x_k)$ bounded
continuous. Ineq.(\ref{Cheb1:eqn})
reduces the problem of showing the convergence of $\{E_{\mu_{M_j}}[F]\}$ to
showing that of $\{E_{\mu_{M_j}}[F; B_R]\}$, where $B_R\equiv
\{\Phi:\|\Phi\|_{{\mathcal H}_s}\leq R\}$.   We write $f_{\leq N}$ as the projection of $f$ onto the
subspace of Fourier modes with $|n|\leq N$ and $f_{> N}$ for
the tail of its series. Now for $\Phi\in B_R$ and for
any $\delta>0$, $|\langle f_j,\Phi\rangle-\langle f_{j,\leq
N},\Phi\rangle |\leq R\|f_{j,>N}\|_{{\mathcal H}_{-s}}<\delta$ for $N$
sufficiently large.  The uniform continuity of $F$ on the bounded set
$\{x: |x_j|\leq R\}$ then gives the convergence.

That $\mu$ is countably additive on the Borel sets generated
by the cylinder sets is also a consequence of 
Ineq.(\ref{Cheb1:eqn}) (\cite{GV}). This concludes the proof
of the proposition.\qed

\appendix
\section{Perturbation theory for the eigenfunctions
of the operator $\mA$}

This appendix summarizes properties of the left eigenfunctions $\{f^{\pm}_{n,\sigma}\}$
 and right eigenfunctions $\{e^{\pm}_{n,\sigma}\}$
and their corresponding eigenvalues $\{\lambda^{\pm}_{n,\sigma}\}$ for the matrix
operator $\mA$.  All estimates are uniform with respect to the ultraviolet cutoff  $M$, and so
the index $M$ is suppressed.  We assume throughout this appendix that the coupling
 functions $\alpha_{1}$ and $\alpha_{2}$
satisfy the assumptions (\ref{alphaAssumption}). See \cite{Kato} for 
the perturbation theory methods utilized.

Fixing the mode number $n\geq 0$ and the ${\pm}$'s and then suppressing these and other indices except as needed, we write
$f^{\pm}_{n,\sigma}= (f_{\sigma,\phi},f_{\sigma,\pi},f_{\sigma,r_1},
f_{\sigma,r_2})$ for the components of $f^{\pm}_{n,\sigma}$, and we write $e^{\pm}_{n,\sigma}=
(e_{\sigma,\phi},e_{\sigma,\pi},e_{\sigma,r_1},f_{\sigma,r_2})^T$ for
the column components of the right eigenfunction.  The corresponding
eigenvalue $\lambda^{\pm}_{n,\sigma}$ is near $\pm i\, \sqrt{n^2
  +1}$; for $n>0$ and for a given choice of $\pm$, the eigenvalue is nearly doubly degenerate, whence
the index $\sigma= 1\,\, {\rm or}\,\,-1$.
 
   From the eigenvalue equation $\langle f^{\pm}_{n,\sigma}\,,\mA\,\,\cdot\rangle_{\mathcal H}= \lambda^{\pm}_{n,\sigma}\langle f^{\pm}_{n,\sigma}\,,\,\,\cdot\rangle_{\mathcal H}$, one finds the relations
\begin{equation}
    f_{\sigma,\phi}= \frac{1}{\lambda_{\sigma}}f_{\sigma,\pi}(\partial_x^2-1), 
    \,\,\,    f_{\sigma,r}= \frac{-\langle f_{\sigma,\pi},\alpha\rangle}{1+\lambda_{\sigma}},
 \end{equation}
and similarly, from $\mA e^{\pm}_{n,\sigma}= \lambda^{\pm}_{n,\sigma}e^{\pm}_{n,\sigma}$,
\begin{equation}\label{esigmaapp.eq}
    e_{\sigma,\phi}= \frac{1}{\lambda_{\sigma}}e_{\sigma,\pi}, \,\,\,
    e_{\sigma,r}= \frac{\langle \alpha, e_{\sigma,\pi}\rangle}{1+\lambda_{\sigma}}.
 \end{equation}
Note that $f_{\sigma,r}$ and $e_{\sigma,r}$ each have two components, e.g., $f_{\sigma,r}= (f_{\sigma, r_1}, f_{\sigma,r_2})$, corresponding to the two components
of $\alpha$ in the above relations. One can then use these relations to write the eigenvalue equation
as a non-linear eigenvalue equation just involving the $\pi$-components,
\begin{equation}
  \lambda_{\sigma}^{2}f_{\sigma,\pi}= f_{\sigma,\pi}(\partial_x^2-1)- \frac{\lambda_{\sigma}}{1+\lambda_{\sigma}}   \langle f_{\sigma,\pi},\alpha\rangle \langle \alpha|
\end{equation}
or 
\begin{equation}\label{eigen.app.eq7} 
 \lambda_{\sigma}^2e_{\sigma,\pi}= (\partial_x^2-1)e_{\sigma,\pi}- \frac{\lambda_{\sigma}}{1+\lambda_{\sigma}} |\alpha\rangle \langle\alpha, e_{\sigma,\pi}\rangle.
\end{equation}
In these equations, $|\alpha\rangle\langle\alpha|= \sum_{i=1}^2|\alpha_i\rangle\langle\alpha_i|$.
For large $n$, $\lambda_{\sigma}/(1+\lambda_{\sigma})= 1\pm i/n +{\mathcal O}(n^{-2})$, so that these latter two equations are nearly self-adjoint eigenvalue
equations for 
    \begin{equation}\label{a1.eq} 
 \lambda_{\sigma}^2e_{\sigma,\pi}= (\partial_x^2-1)e_{\sigma,\pi}- \beta \alpha\langle\alpha, e_{\sigma,\pi}\rangle,
\end{equation}
with $\lambda_{\sigma}$ determined implicitly by 
substituting in $\beta = \lambda_{\sigma}/(1+\lambda_{\sigma})$. Evidently,
$e_{\sigma,\pi}= -\beta (\lambda_{\sigma}^{2}-\partial_x^2+1)^{-1}
\alpha \langle\alpha, e_{\sigma,\pi}\rangle$, and from this representation, one sees 
that the $m^{th}$ Fourier coefficient $\hat{e}_{\sigma,\pi}(m)$ of $e_{\sigma,\pi}$,   is ${\mathcal O}(\frac{n^{\theta}}{n^{2}-m^{2}})$, for
$m^{2}\neq m^{2}$. This implies in particular  that
\begin{equation}\label{Aae.ineq}
             \langle \alpha,  e_{\sigma,\pi}\rangle= {\mathcal O}(n^{\theta}).
\end{equation}

In the following, let $P_0= P_0(n)$ be the projection onto the eigenspace spanned
by $e^{inx}$ and  $e^{-inx}$ in $L^2[0,2\pi]$.
\begin{lemma}\label{appendixlemma}
   We have that for $n$ large,
 the eigenvalues $\{\lambda_{n\sigma}^{\pm}\}$
of $A$ are given by
\begin{eqnarray}\label{eigenshift5appendix.eqn10}
  {\lambda_{n\sigma}^{\pm}}&=& \pm \left(i\sqrt{n^2+1} +\frac{(1\mp i/n)\mu_{n,\sigma}}{2i n}\right)\\
&&\phantom{XX} +i{\mathcal O}(n^{4\theta-2}\ln{n})
 +{\mathcal O}(n^{4\theta-3}\ln{n})\nonumber
\end{eqnarray} 
where the $\mu_{n,\sigma}$ are the two eigenvalues of the 
operator $-P_0(n)\alpha\rangle\langle \alpha P_0(n)$,  and 
where the error terms are, respectively, imaginary and real.

Let $e^{\pm}_{n,\sigma,r}$ be the $r$-components of the right eigenvector
 $e^{\pm}_{n,\sigma}$ of $A$.  Then for $\sigma \neq \sigma'$, we have 
 \begin{equation}\label{Vorthog.appendix.eq11}
   \frac{ e_{n,\sigma,r}^{\pm\,\,*}\cdot  e_{n,\sigma',r}^{\pm}}{|e_{n,\sigma,r}^{\pm}|\,|e_{n,\sigma',r}^{\pm}|}=
       {\mathcal O}(n^{2\theta-1}\ln{ n}),
\end{equation}
where the dot indicates the dot product of the components, $|\cdot|$ being the Euclidean
length. 
\end{lemma}

\noindent {\it Sketch of proof}:
     For $n>0$, let $P= P(n)$ be the projection onto the subspace spanned
by the two eigenvectors corresponding to the two eigenvalues $\lambda_{\sigma}^2$
near $-(n^2+1)$ for the operator $\partial_x^2-1-\beta |\alpha\rangle \langle \alpha|$.
 ($P$ is close to $P_0$.) Then the shift in the eigenvalues is determined from the $2\times 2$ matrix equation 
\begin{equation}
      (\lambda_{\sigma}^2+ (n^2+1))P_0PP_0 \xi_{\sigma}= -\beta P_0\alpha\rangle\langle \alpha PP_0 \xi_{\sigma}
\end{equation}
for $\xi_{\sigma}$ in the span of $P_0$.  The projection $P$ itself can be estimated
from its representation as a contour integral of the resolvent
$(\partial_x^2-1-\beta|\alpha \rangle\langle \alpha|-z)^{-1}$ expanded
in a Neumann series. One finds that
 \begin{equation}\label{muest.app.eqn4}
   \lambda_{\sigma}^2= -(n^2+1)+\beta\mu_{\sigma}+\beta^2{\mathcal O}(n^{4\theta-1}\ln{ n}),
\end{equation}
where $\mu_{\sigma}$ is one of the two eigenvalues of the rank $2$ matrix
$-P_0\alpha\rangle\langle \alpha P_0$ and is of the order
of $|\hat{\alpha}(n)|^2\equiv\sup|\hat{\alpha_i}(n)|^2={\mathcal O}(n^{2\theta})$.
The Neumann series for the resolvent above
is in powers of $\langle \alpha, (\partial_x^2-1-z)^{-1}\alpha\rangle$ with $z$ traversing a circle $\{z:\,|z+ n^2+1|= n/2\}$. These
powers are estimated using
\begin{eqnarray}
\lefteqn{|\langle \alpha, (\partial_x^2-1-z)^{-1}\alpha\rangle|}&&\nonumber\\
&\leq& c\!\!\! \!\!\!\sum_{\{m : m\leq n/2\}}\frac{|\hat{\alpha}(m)|^2 }{n^2} + \,\,\,\,
c\!\!\!\!\!\!\!\!\!\!\!\! \sum_{\{m:  n/2\leq m\leq 3n/2, m\neq n\}}\frac{|\hat{\alpha}(n)|^2 }{|n^2-m^2|}\nonumber\\
&& +c\frac{|\hat{\alpha}(n)|^2 }{n} +c\!\!\!\!\!\!  \sum_{\{m : m\geq 3 n/2\}}\frac{|\hat{\alpha}(m)|^2 }{m^2}
\nonumber\\
&\leq & \frac{c |\hat{\alpha}(n)|^2\ln{n}}{n}.
\end{eqnarray}
(The second sum in the second line accounts for the $\ln{n}/n$ factor, by an integral test.)
One uses this estimate to obtain the correction to the eigenvalue shift
in Eq.(\ref{muest.app.eqn4}). See \cite{TW} for additional details.
We then set $\beta=\lambda_\sigma/(1+\lambda_{\sigma})$ into Eq.(\ref{muest.app.eqn4}) to determine
$\lambda_{\sigma}$ implicitly; this gives the first assertion of the lemma,
Eq.(\ref{eigenshift5appendix.eqn10}).

    We also need an approximate orthogonality relation for $e_{\sigma,r}$
and $e_{\sigma',r}$, $\sigma\neq \sigma'$.  In  Eq.(\ref{eigen.app.eq7}),  we decompose $e_{\sigma,\pi}$ as
\begin{equation}\label{decom.appendix.eqq}
e_{\sigma,\pi}= \xi_{\sigma}^0 + \tilde{e}_{\sigma,\pi},
\end{equation}
 where $\xi_{\sigma}^0$ is an $L^2$-normalized eigenfunction of $P_0\alpha\rangle\langle \alpha P_0$, and hence in the subspace spanned
by $P_0$.   Then  
\begin{equation}\label{aux.appendix.eq}
   P_0(\lambda_{\sigma}^2+n^2+1-\beta\alpha\rangle\langle\alpha|)
(\xi_{\sigma}^0 + P_0\tilde{e}_{\sigma,\pi}+Q_0\tilde{e}_{\sigma,\pi})= 0,
\end{equation}
where $Q_0= 1-P_0$. We have that
\begin{equation}\label{aux.appendix2.eq}
    \|P_0(\lambda_{\sigma}^2+n^2+1-\beta\alpha\rangle\langle\alpha|)\xi_{\sigma}^0\|= {\mathcal O}(n^{4\theta-1}\ln{ n}) 
\end{equation}
by the eigenvalue shift estimate above, Ineq.(\ref{muest.app.eqn4}). Also,
we have that
\begin{equation}\label{motherappendix.eq1}
  Q_0\tilde{e}_{\sigma,\pi} = -\beta Q_0(\lambda_{\sigma}^2-\partial_x^2+1)^{-1}\alpha\rangle\langle\alpha, e_{\sigma,\pi}\rangle,
\end{equation}
so 
\begin{equation}\label{Qappendix.eq3}
   |\langle\alpha Q_0\tilde{e}_{\sigma,\pi}\rangle|= {\mathcal O}(n^{3\theta-1}\ln{ n}), 
\end{equation}
and thus
\begin{equation}
 \|P_0(\lambda_{\sigma}^2+n^2+1-\beta\alpha\rangle\langle\alpha|)Q_0\tilde{e}_{\sigma,\pi})\|= {\mathcal O}(n^{4\theta-1}\ln{ n}). 
\end{equation}
 From the identity Eq.(\ref{aux.appendix.eq}), this last equation,
and Eq.(\ref{aux.appendix2.eq}) above, it follows that
\begin{equation}
   \left(\mu_{\sigma}-\mu_{\sigma'}+{\mathcal O}(n^{4\theta-1}\ln{ n})\right)\| P_0\tilde{e}_{\sigma,\pi}\|= {\mathcal O}(n^{4\theta-1}\ln{ n}),
\end{equation}
where $\sigma'$ refers to the complementary value of $\sigma$.
The assumption on the coupling functions $\alpha$, Eq.(\ref{alphaAssumption}),  assures
that $|\mu_{\sigma}-\mu_{\sigma'}|\geq c |\hat{\alpha}(n)|^2$, so that
\begin{equation}\label{Poappendix.eqn9}
       \| P_0\tilde{e}_{\sigma,\pi}\|=  {\mathcal O}(n^{2\theta-1}\ln{ n}).
\end{equation}

The last equation of the lemma, Eq.(\ref{Vorthog.appendix.eq11}), follows
from the equation for $e_{\sigma,r}$, Eq.(\ref{esigmaapp.eq}); the decomposition for $e_{\sigma,\pi}$ in Eq.(\ref{decom.appendix.eqq}); and  the orthogonality of $\xi_{\sigma}^0$ and  $\xi_{\sigma'}^0$
under $P_0\alpha\rangle\langle\alpha P_0$. We have that
\begin{eqnarray}
   \lefteqn{\langle e_{\sigma,\pi},\alpha\rangle\langle \alpha,e_{\sigma',\pi}\rangle}&&\nonumber\\
&=& \langle \xi_{\sigma}^0, \alpha\rangle\langle\alpha, Q_0\tilde{e}_{\sigma',\pi}\rangle +\langle Q_0\tilde{e}_{\sigma,\pi},\alpha\rangle\langle \alpha,\xi_{\sigma',\pi}^0\rangle +  \langle P_0\tilde{e}_{\sigma,\pi},\alpha\rangle \langle\alpha,Q_0 \tilde{e}_{\sigma',\pi}\rangle  \nonumber\\
&& +  \langle Q_0\tilde{e}_{\sigma,\pi},\alpha\rangle \langle\alpha,P_0 \tilde{e}_{\sigma',\pi}\rangle  + \langle Q_0\tilde{e}_{\sigma,\pi},\alpha\rangle \langle\alpha,Q_0 \tilde{e}_{\sigma',\pi}\rangle        \nonumber\\
&=& {\mathcal O}(n^{4\theta-1}\ln{ n})
\end{eqnarray}
by Eqs.(\ref{Qappendix.eq3},\ref{Poappendix.eqn9}).  Combining
this equation with Eq.(\ref{esigmaapp.eq}), we obtain Eq.(\ref{Vorthog.appendix.eq11}) of the lemma  \qed

Eq.(\ref{ineqtable.eq}) of the text provides estimates on inner products $ \langle f^{\pm}_{n,\sigma}(0),e^{\pm'}_{m,\sigma'}\rangle_{\mathcal H}$. To illustrate how
these estimates are obtained, consider a case $m\neq n$, $m,n$ large. By familiar
resolvent identities, and with $\mB$ the unperturbed matrix operator with the
$\alpha$'s turned off and with the left eigenfunction $f^{\pm}_{n,\sigma}(0)$,
\begin{eqnarray}\label{f0eappendix.eq10}
  \lefteqn{\langle f^{\pm}_{n,\sigma}(0),e^{\pm'}_{m,\sigma'}\rangle_{\mathcal H}=
   \langle f^{\pm}_{n,\sigma}(0),P_0(n)(P(m)-P_0(m))e^{\pm'}_{m,\sigma'}\rangle_{\mathcal H}}&&\nonumber\\
&=& -\frac{1}{2\pi i}\int_{\gamma_m} dz
\langle f^{\pm}_{n,\sigma}(0),\frac{1}{\mB-z}\mC\frac{1}{\mA -z}e^{\pm'}_{m,\sigma'}\rangle_{\mathcal H}\nonumber\\
&=&  \frac{1}{2\pi i}\int_{\gamma_m} dz \frac{\langle f^{\pm}_{n,\sigma,\pi}(0),\alpha\rangle e^{\pm'}_{m,\sigma',r}}{(\lambda_n^{\pm}(0)-z)(\lambda_{m,\sigma'}^{\pm'}-z)}\nonumber\\
&\sim& \frac{\hat{\alpha}(n) \hat{\alpha}(m)}{(\lambda_n^{\pm}(0)-\lambda_{M,m,\sigma'}^{\pm'})(\lambda_{M,m,\sigma'}^{\pm'}+1)}= {\mathcal O}\left(\frac{n^{\theta}m^{\theta-1}}{n-m}  \right).
\end{eqnarray}
Here, $\gamma_m$ is the circle $\{z: |z-\lambda_{m,\sigma'}^{\pm'}|= 1/2\}$ and, in the second line, $\mC=\mA-\mB$ is the $4\times 4$
 perturbation matrix of just the $\alpha$ entries, all other entries being zero.  In the last line
we have used Eq.(\ref{eigen.app.eq7}) for $e^{\pm'}_{m,\sigma',r}$ and
the fact that $\langle \alpha,e^{\pm'}_{m,\sigma',\pi}\rangle\sim \hat{\alpha}(n)=
{\mathcal O}(n^{\theta})$, Eq.(\ref{Aae.ineq}). The other relations of Eq.(\ref{ineqtable.eq}) are analyzed similarly.

\acknowledgements This article is based on the Ph.D. thesis of YW at 
the University of Virginia.  
\endacknowledgements

\end{document}